\documentclass[12pt]{article} 
\usepackage{graphicx}  
\begin{document} 
\begin{center} 
{\bf Persistence in the zero-temperature dynamics of the $Q$-states Potts model on undirected-directed Barab\'asi-Albert networks and  Erd\"os-R\'enyi random graphs }$^{\star}$ 
\end{center}  
 
\bigskip 
\centerline{F. P. Fernandes and F.W.S. Lima } 
  
\bigskip 
\noindent 
Departamento de F\'{\i}sica,  
Universidade Federal do Piau\'{\i}, 64049-550, Teresina - PI, Brazil \\ 
 
\medskip 
  e-mail:  fwslima@gmail.com 
\bigskip 
 
$\star$ This paper is dedicated to Dietrich Stauffer for his great contribution to physics. 
 
 
{\small Abstract: The zero-temperature Glauber dynamics is used to investigate the persistence probability $P(t)$ in the Potts model with $Q=3,4,5,7,9,12,24,64 ,128$, $256, 512, 1024,4096,16384 $,..., $2^{30}$  states on {\it directed} and {\it undirected} Barab\'asi-Albert networks and  Erd\"os-R\'enyi random graphs. 
In this model it is found that $P(t)$ decays exponentially to zero in short times for {\it directed} and {\it undirected} Erd\"os-R\'enyi random graphs. For {\it directed} and {\it undirected} Barab\'asi-Albert networks, in contrast it decays exponentially to a  constant value for long times, i.e, $P(\infty)$ is different from zero for all $Q$ values (here studied) from $Q=3,4,5,..., 2^{30}$; this shows "blocking" for all these $Q$ values. Except that for $Q=2^{30}$ in the {\it undirected} case $P(t)$ tends exponentially to zero; this could be just a finite-size effect since in the other "blocking" cases you may have only a few unchanged spins.  
} 
\bigskip

Keywords: Monte Carlo simulation, spins , networks, Potts. 
  
\bigskip 
 {\bf Introduction}

 Derrida et al. \cite{derrida1} discovered that the fraction $P(t)$ of spins which have not flipped until time $t$ exhibits non-trivial dynamical behavior for the Potts model and zero-temperature dynamics. In one dimension, they showed that $P(t)$ decays algebraically, with an exponent $\theta$ varying continuously between $0$ and $1$ as $Q$ increases from 1 to $\infty$ \cite{derrida1,derrida2}. Derrida et al. \cite{derrida3} studied the problem in two dimensions by Monte Carlo simulations. However, their result for high $Q$ was inaccurate since the data showed some curvature and only effective exponents for $\theta$  could be determined. Michael Hennecke \cite{MH1} investigated the two-dimensional Potts model for $Q=\infty$ by a similar zero-temperature Monte Carlo method. Due to a much higher numerical effort, its has been confirmed that asymptotically, $\theta$ approaches unity if $Q=\infty$. Michael Hennecke \cite{MH2} presented also simulation results for finite $Q$  
in the order to obtain precise, and hopefully asymptotic, values for the exponents $\theta$ for general $Q$. He applied the Monte Carlo Method described in Ref. \cite{MH1} to  a triangular lattice with nearest-neighbor interactions and linear dimension $L=2000$, where a lattice site is chosen at random and the trial spin is set to the color of the majority of neighbors. For the pure ferromagnetic one-dimensional Ising model, persistence was studied numerically \cite{derrida1}, and for the one-dimensional Potts model the exponent was analytically computed \cite{derrida2}. For higher dimensions evidence for power law decay was obtained numerically \cite{stauffer}, while the exponent was analytically obtained \cite{maj,derrida4} showing that $P(t)$ decays algebraically as 
 
\begin{equation} 
P(t)\propto t^{-\theta(d)}, 
\end{equation} 
where $\theta(d)$ is a a non-trivial persistence exponent. For the same ferromagnetic Ising model at higher dimensions \cite{stauffer} $(d>4)$ and for the two-dimensional ferromagnetic $Q$-states Potts model \cite{derrida5} $(Q>4)$ it has been shown through computer simulations that $P(t)$ is not equal to zero for $t\to  \infty$. This characteristic is sometimes called "blocking". In this way, if  $P(\infty)>0$ we can reformulate the problem by restricting our observation to those spins that eventually flip. Hence, we can consider the behavior of the residual persistence given by 
\begin{equation} 
p(t)=P(t)-P(\infty). 
\end{equation} 
Recently attention has been turned to the persistence problem in systems containing disorder \cite{jain,cm}. Numerical simulations of the zero-temperature dynamics of bond diluted (weak dilution \cite{jain} or strong dilution \cite{jain1}) two-dimensional Ising model also reported "blocking" evidence. Howard has found evidence of an exponential decay of the persistence with "blocking" for the homogeneous tree of degree three(T) with random spin configuration at the initial time \cite{howard}. In order to investigate other aspects of the disorder effect to the persistence problem, Lima et al. \cite{lima} presented new data for the zero-temperature dynamics of the two-dimensional ferromagnetic Ising model on a random Poissonian lattice. They showed that this system exhibits "blocking" which means that the persistence does not go to zero when $t\to \infty$. They showed also that $p(t)$ decays exponentially in the long-time regime.  Recently S. Jain et al. \cite{jain2}  
studied the persistence problem in a random bond Ising model of socio-econo dynamics. In this study they found no evidence of the "blocking" effect. Here in this paper we study the persistence problem in the Potts model with $Q$ states ($Q=3,4,...,2^{30}$) on {\it directed} and {\it undirected} Barab\'asi-Albert(BA) \cite{ba} networks and Erd\"os-R\'enyi(ER) \cite{erdo} random graphs using the zero-temperature Glauber dynamics. The results show that for {\it directed} and {\it undirected} ER random graphs the persistence decays exponentially to zero without evidence of "blocking", but for {\it directed} and {\it undirected} BA networks the persistence decays exponentially with and without presence of the "blocking". In both cases no algebraic decay of the persistence with the time was observed. 
 
{\bf Model and simulation} 
  
We consider the $Q$-states Potts model on {\it directed} and {\it undirected}     Barab\'asi-Albert(BA) networks and Erd\"os-R\'enyi(ER) random graphs defined by the Hamiltonian 
\begin{equation} 
H=-J\sum_{<ij>}\delta_{S_{i},S_{j}} 
\end{equation} 
here $J>0$ is a ferromagnetic coupling, the sum is over all neighboring interactions included in the system, the spin values are in the range $S_{i}=1,...,Q$, and $\delta_{x,y}$ is Kronecker's delta function. We perform simulation over networks with $N=10^{6}$ sites. A random initial configuration of spins is obtained and $P(t)$ is calculated over  $10^{5}$ and $10^{6}$ Monte Carlo steps and a quenched average is done over $10$ different networks for each Monte Carlo step. The zero-temperature Glauber dynamics was used in order to check the number of spins that never changed their state up to time $t$. In this dynamics, we start with a random initial spin configuration and allow it to be flipped at random or following a given logical sequence. The selected spin, $S_{i}$, is flipped or not according to $\Delta E_{i}$, where $\Delta E_{i}$ is the energy variation of the site $i$. If $\Delta E_{i}<0$ the spin $S_{i}$ is flipped with probability one. 
If $\Delta E_{i}=0$ the spin $S_{i}$ is flipped with probability $1/2$ chosen at random. Finally, if $\Delta E_{i}>0$ the spin is not flipped. One Monte Carlo step corresponds to application of the above rule for all spins of the network. The system configuration is left to evolve until a given Monte Carlo step $10^{5}$ and $10^{6}$. The number $n(t)$ of spins that did not change up to this time $t$ is computed for each Monte Carlo step for determination of the persistence probability given by \cite{derrida1,jain,jain1} 
\begin{equation} 
P(t)=\frac{<n(t)>}{N}. 
\end{equation} 
where $n(t)$ is the sites number that did not flip their state up to time $t$ and $N$ is the number of sites. 
\begin{figure}[hbt] 
\begin{center} 
\includegraphics [angle=0,scale=0.60]{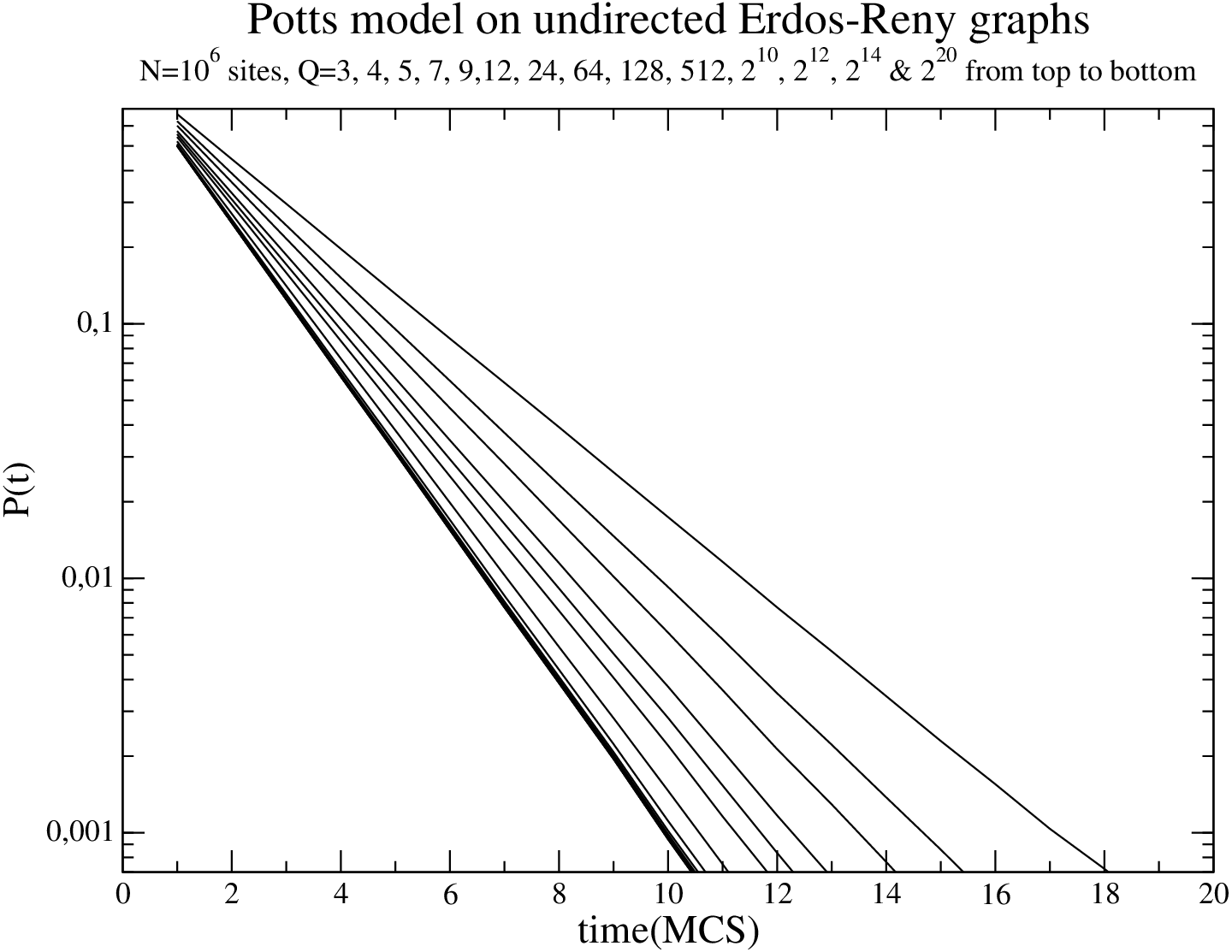} 
\end{center} 
\caption{$P(t)$ (on logarithmic scale) versus time for various $Q$ and $ m=2$ on {\it undirected} Erd\"os-R\'enyi(ER) random graphs. $Q=3,4,5,...,2^{20}$ from top to bottom.} 
\end{figure} 
\bigskip 
 
\begin{figure}[hbt] 
\begin{center} 
\includegraphics [angle=0,scale=0.60]{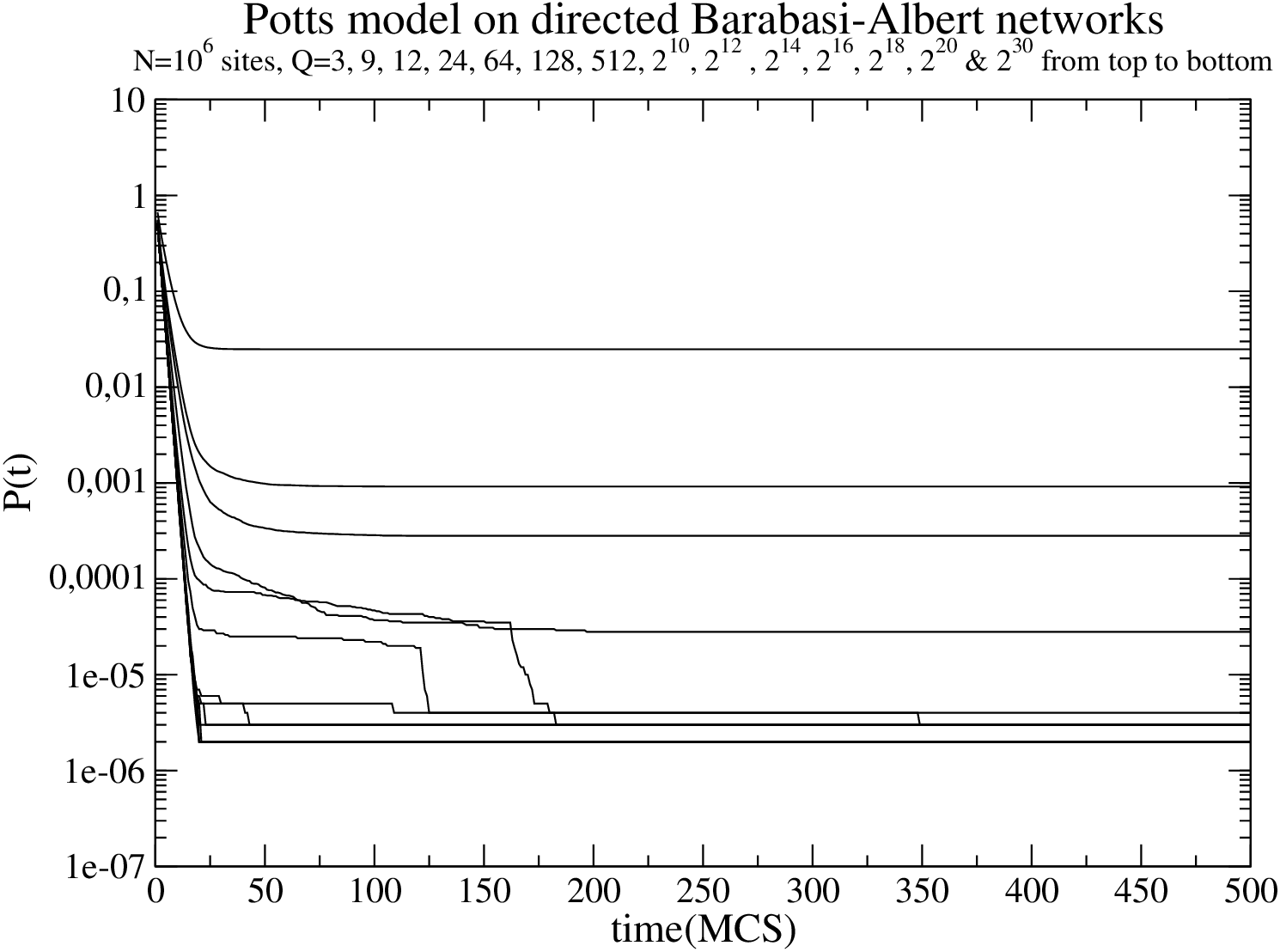} 
\end{center} 
\caption{$P(t)$ (on logarithmic scale) versus $time$ for various $Q $ and $ m=2$, on  {\it directed} Barab\'asi-Albert(BA) networks. $Q=3,4,5,...,2^{30}$ from top to bottom.} 
\end{figure}  
\bigskip 
\begin{figure}[hbt] 
\begin{center} 
\includegraphics [angle=0,scale=0.60]{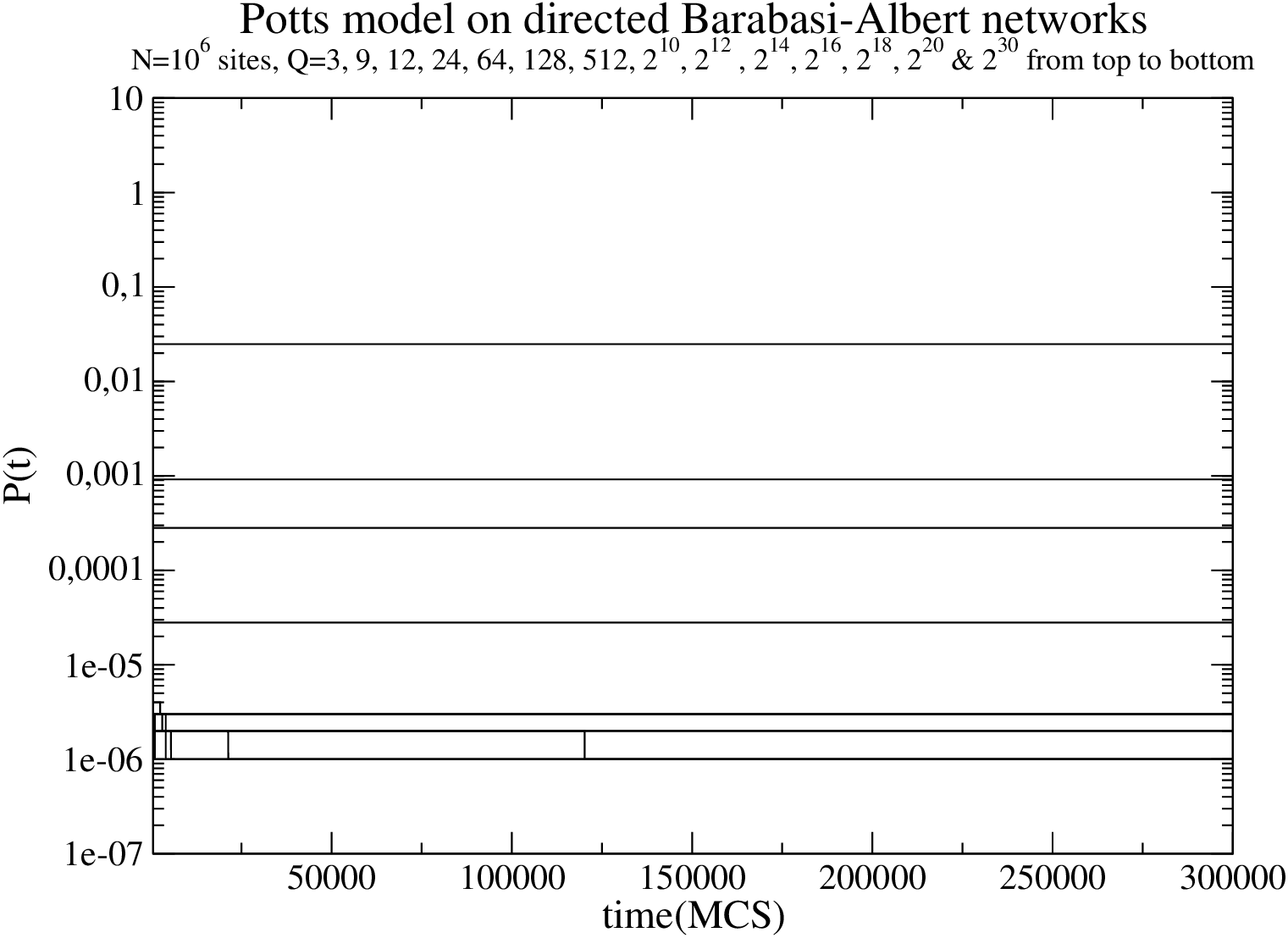} 
\end{center} 
\caption{$P(t)$ (on logarithmic scale) versus $time$ for various $Q $ and $ m=2$, on  {\it directed} Barab\'asi-Albert(BA) networks. $Q=3,4,5,...,2^{30}$ from top to bottom. The time varies from 500 to 300000.} 
\end{figure}  
\bigskip 
\begin{figure}[hbt] 
\begin{center} 
\includegraphics [angle=0,scale=0.60]{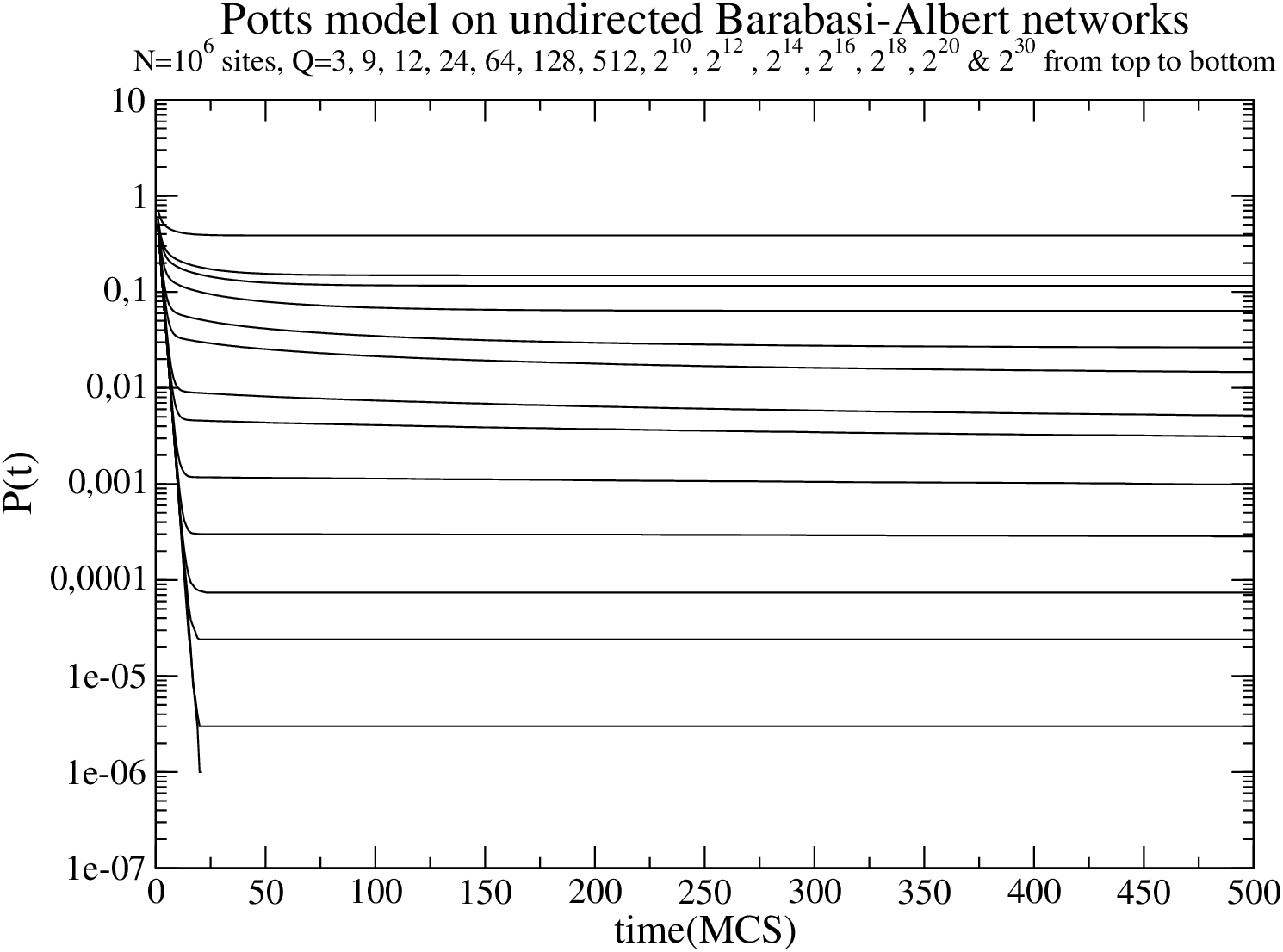} 
\end{center} 
\caption{$P(t)$ (on logarithmic scale) versus $time$ for various $Q $ and $ m=2$ on  {\it undirected} Barab\'asi-Albert(BA) networks. $Q=3,4,5,...,2^{30}$ from top to bottom. } 
\end{figure}  
\bigskip 
\begin{figure}[hbt] 
\begin{center} 
\includegraphics [angle=0,scale=0.60]{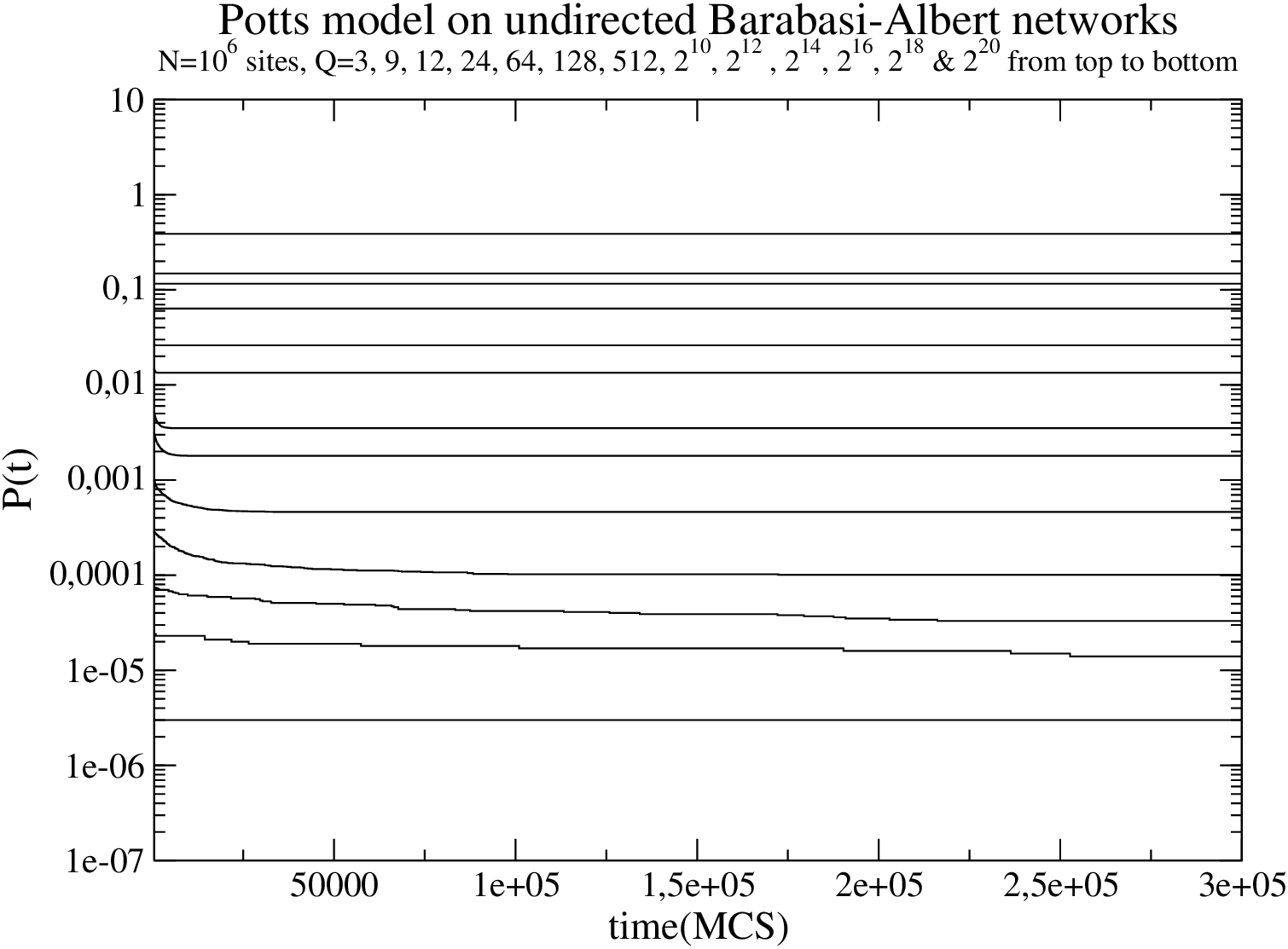} 
\end{center} 
\caption{$P(t)$ (on logarithmic scale) versus $time$ for various $Q $ and $ m=2$ on  {\it undirected} Barab\'asi-Albert(BA) networks. $Q=3,4,5,...,2^{22}$ (because $Q=2^{30}$ is zero in this interval)  
from top to bottom. The time of $500$ to $300000$. } 
\end{figure}  
\bigskip 
\begin{figure}[hbt] 
\begin{center} 
\includegraphics [angle=0,scale=0.60]{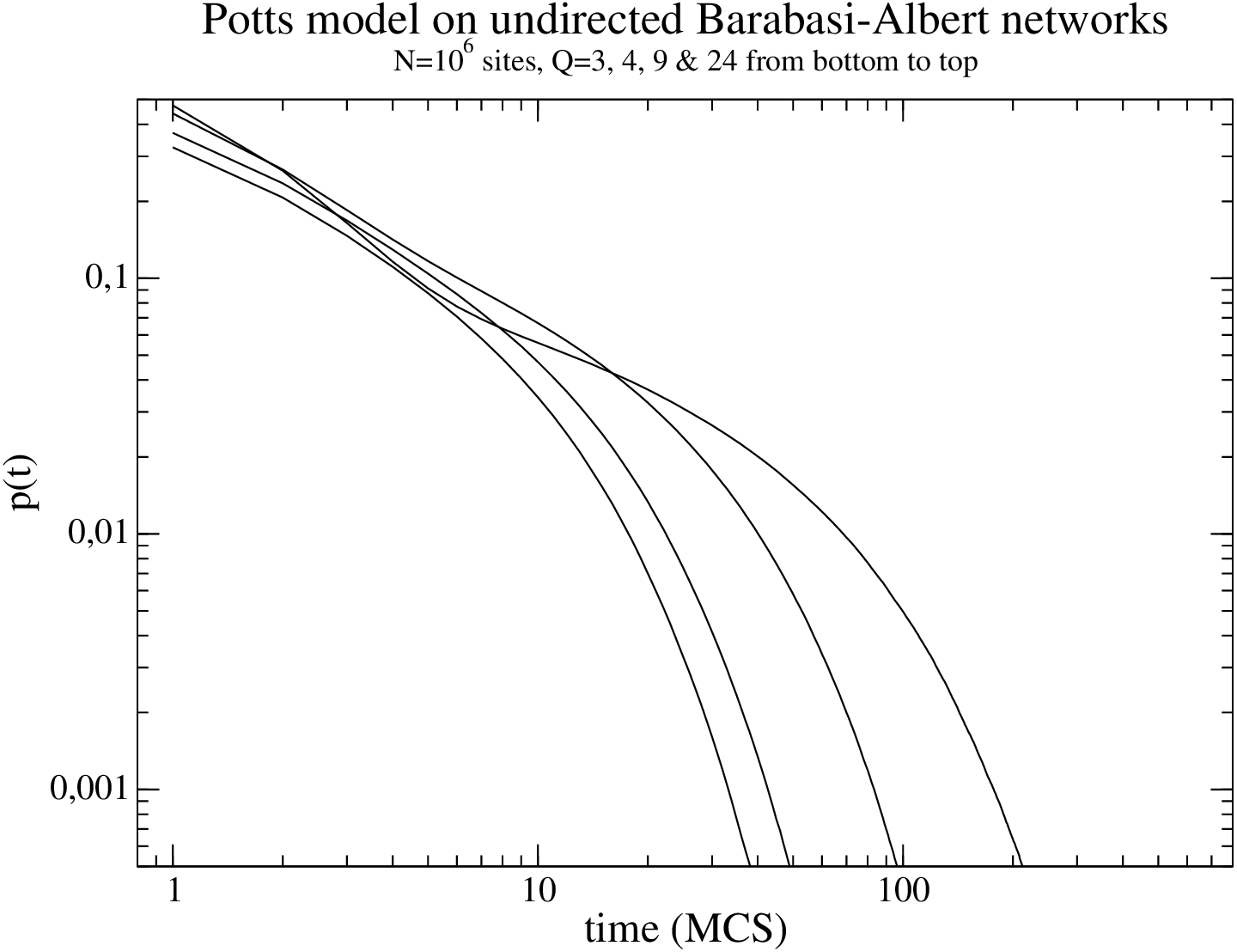} 
\end{center} 
\caption{$p(t)$ versus time (double-logarithmic scales) for various $Q=3,4,9, 24 $ (from bottom to top) and $ m=2$ on  {\it undirected} Barab\'asi-Albert(BA) networks.  } 
\end{figure}  
\bigskip 
\begin{figure}[hbt] 
\begin{center} 
\includegraphics [angle=0,scale=0.60]{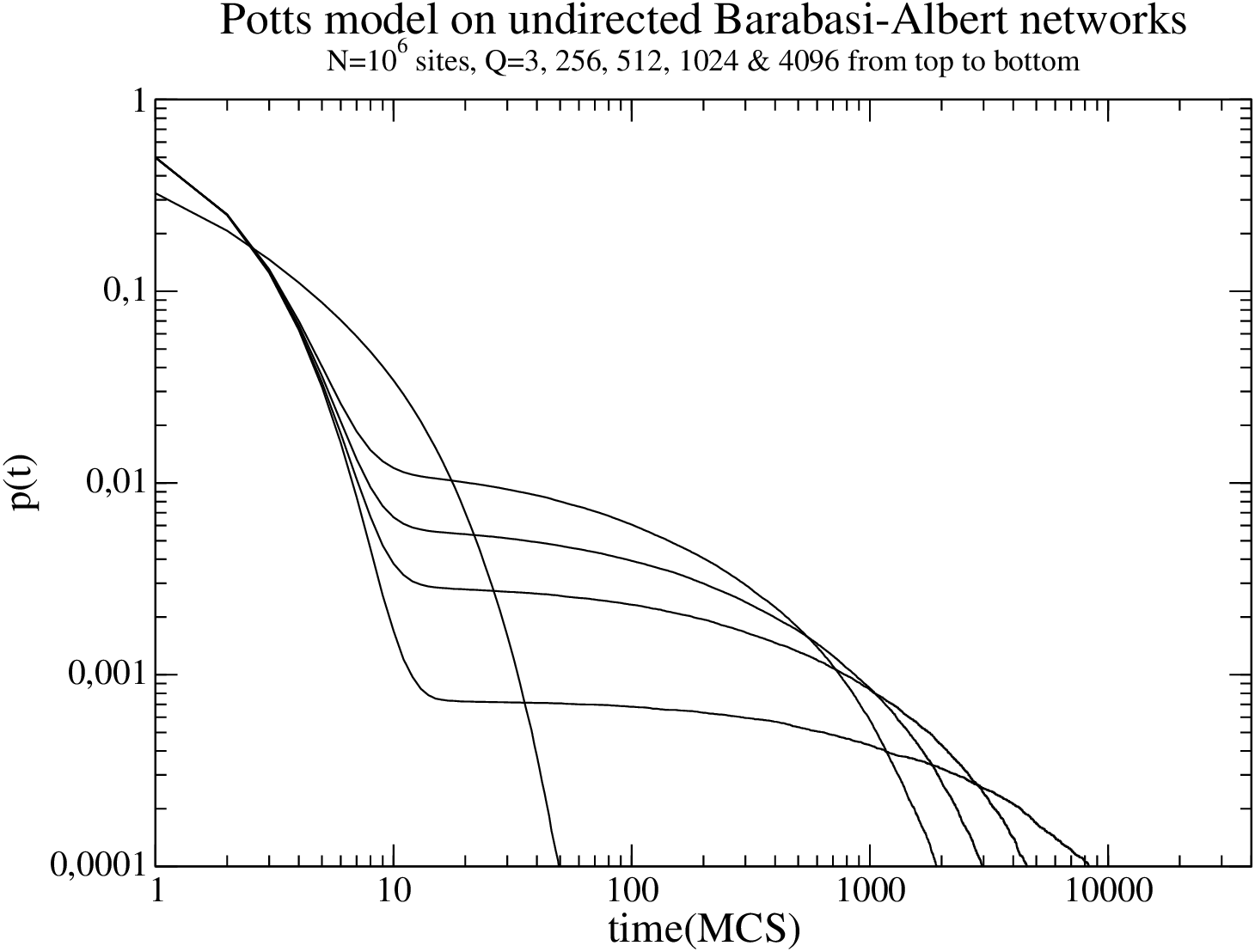} 
\end{center} 
\caption{$p(t)$ versus time (double-logarithmic scales) for $Q=3$ and other various values of  $Q$ $(256,512,1024,4096)$ (from top to bottom) and $ m=2$ on  {\it undirected} Barab\'asi-Albert(BA) networks.  } 
\end{figure}  
\bigskip 
\begin{figure}[hbt] 
\begin{center} 
\includegraphics [angle=0,scale=0.60]{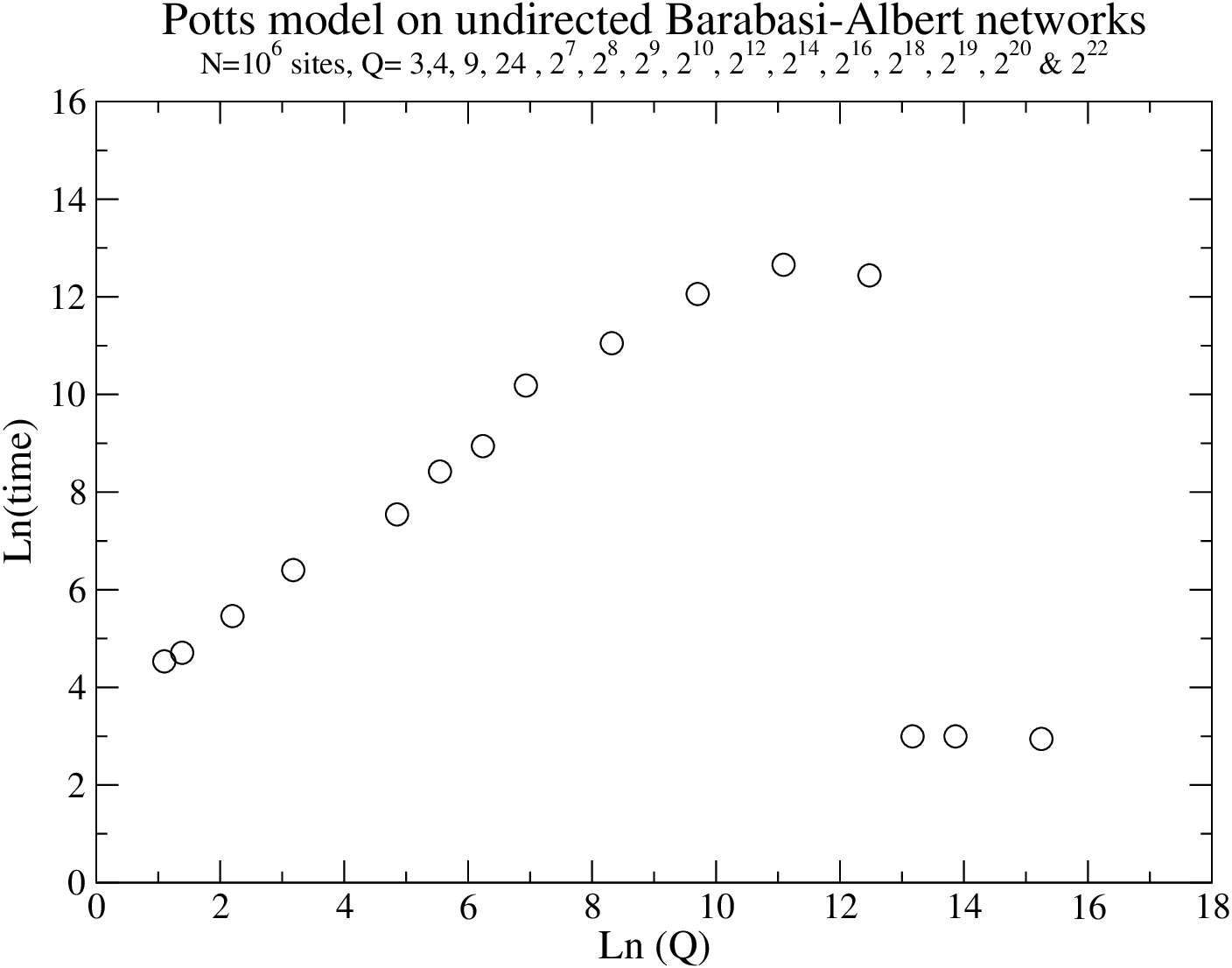} 
\end{center} 
\caption{$\ln(time)$ versus $ln(Q)$ for various $Q=3, 4,5,...,2^{22} $ and $ m=2$ on  {\it undirected} Barab\'asi-Albert(BA) networks, where $time$ is the number of time steps before "blocking" occurs. Here we observe that $time$  increase when $Q$ increases for $Q=3,4,5,...,2^{14}$. The best linear fit gives a slope=0.89(3). Afterwards the time decays dramatically to short times for $Q=2^{19}, 2^{20}, 2^{22}$. } 
\end{figure} 
{\bf Results and Conclusion} 
 
In Fig. 1 we plot $P(t)$ versus $t$ for Potts models with $Q=3,4,5,7,9,12,$ $24,64 ,128$, $256, 512, 1024,4096,16384 $ and $2^{20}$ states in $N=10^{6}$ sites and $m=2$ (coordination  probability $p=m/N$ for ER random graphs) on  {\it undirected} ER random graphs. The persistence probability $P(t)$ seems to decay
 exponentially to zero for all $Q$ values. For {\it directed} ER random graphs
the results are the same. In Figs. 2 and 4 we plot again $\log P(t)$ versus $t$ (for $t$ from $1$ to $500$) for Potts model for same $Q$ values ($Q=3,4,...,2^{30}$) and $m=2$ (where here $m$ is also the size of the initial core for generating the BA networks) values,  
for {\it directed} and {\it undirected} BA networks, respectively. We observe that for $Q=3,4,5,7,9$,..., $2^{30}$ the persistence probability decays exponentially to a constant value $P(\infty) > 0$ for {\it directed} BA networks. This exhibits the evidence of the "blocking" which means that the persistence does not go to zero when $t \to \infty$. Therefore, for {\it undirected} BA networks the persistence decays exponentially also to a constant value $P(\infty)$ for all finite $Q$ ($Q < 2^{22})$ of the Potts model investigated when $t \to \infty$, again we have evidence the "blocking" for these finite
 $Q$ values, but for $Q=2^{30}$ ("$Q=\infty$") the persistence
 probability decays exponentially to zero. In the Fig. 3 and 5 we plot the same as Fig. 2 and 4, but now for $t$ from $500$ to $3 \times 10^{5}$ for $Q$ values ($Q=3,4,5,7,9$,..., $2^{30}$ and $Q=3,4,5,7,9$,..., $2^{22}$) for {\it directed} and {\it undirected} BA networks, respectively, showing that (in this observed interval) there are no evidences for phase transistions  
from finite values of the persistence probability $P(t\rightarrow \infty)$ to zero for any value of $Q$. For {\it directed} BA networks, $P(t)$ decays to values that correspond to 1, 2 or 3 survivors of $10^{6}$ sites. For {\it undirected} $Q=2^{30}$  states the persistence probability decays quickly to zero, showing that there are survivors for $Q<2^{22}$. 
In Fig. 6 we plot double-logarithmically the behavior of $p(t)$ versus time for some $Q=3,4,9, 24 $ for {\it undirected} BA networks. This
 result showed the exponential relaxation of $p(t)$ $\to 0$. In Fig.7 we plot the same variables, but now for $Q=3, 256, 512, 1024, 4096$.  
In this picture we observe for $Q=3$ an exponential decay to zero, but for $Q=256, 512, 1024, 4096$, we see some kind of 
algebraic decline, followed by an exponential decline for $p(t)$ for longer times. 
In Fig.8 we plot $\ln(time)$ versus $\ln (Q)$ for $Q=3,4,...,2^{22}$ values and observe the $time$ the system needs to reach the "blocking" state on {\it undirected} BA networks. Initially we observe a growth of the $time$ for values from $Q=3$ until $2^{14}$ states, after that the $time$ decreases for large  $Q$-states values. The system needs only a short $time$ to reach the "blocking" state for large  $Q$-states values. The linear best fit gives the slope=$0.89(3)$ for growth of the $time$ for values from $Q=3$ until $2^{14}$ states. 
\bigskip

In summary, we have presented new data for the zero-temperature dynamics of ferromagnetic Potts models on scale free networks and random graphs. For the {\it directed} and {\it undirected} Erd\"os-R\'enyi random graphs the system does not exhibit evidence of "blocking" independently of the assumed values for $Q$, that is, the persistence probability $P(t)$ go to zero when $ t \to \infty$. In contrast, for {\it directed} and {\it undirected} BA networks the system exhibit evidence of "blocking" for $Q=3,4,5,7,9$, 
..., $2^{30}$ values, here observed, when for $t \to \infty$ the persistence probability does not go to zero. 
For $Q$ large the "blocking" there is with only $1,2,3$ or $4$ survivors of the 
initial total of the $10^{6}$ survivors. 
In the {\it undirected} case we observe that the persistence probability $P(t)$ decays exponentially to values different from zero for $Q=3,4,5,7,9,...,2^{22}$ when $t \to \infty$, that shows the evidence the "blocking": but for  $Q= 2^{30}$ this persistence decays to zero, this could be just a finite-size effect since in the other "blocking" cases one may have only few unchanged spins. Here we also study the residual persistence probability $p(t)$ for {\it undirected} BA networks, and believe that for small values $Q=3,4,9, 24$ this decays exponentially to zero, but for $Q=256, 512, 1024, 4096$, we see first some kind of algebraic decay, followed by exponential decline to zero.  
We also observe the increase of the time to reach "blocking" with increasing $Q$, that initially is linear in the log-log scale of the  $Q=3, 4, 7,...,2^{14}$ and afterwards the time decays dramatically to short times for $Q=2^{19}, 2^{20}, 2^{22}$.

   The authors thanks  D. Stauffer for many suggestions and fruitful 
discussions during the development this work and also for the revision of 
this paper. We also acknowledge the Brazilian agency FAPEPI 
(Teresina-Piau\'{\i}-Brasil) for  its financial support. This work was also supported by the 
system SGI Altix 1350 the computational park CENAPAD.UNICAMP-USP, SP-BRAZIL.


\begin{thebibliography}{99} 
 
\bibitem{derrida1} B. Derrida, A. J. Bray, C. Godreche, J. Phys. A 27 (1994) L357.  
  
\bibitem{derrida2} B. Derrida, V. Hakim, V. Pasquier, Phys. Rev. Let. 75 (1995) 751. 
 
\bibitem{derrida3}B. Derrida, P. M. C. de Oliveira, D. Stauffer, Physica A 244 (1996) 604.  
 
\bibitem{MH1} M. Hennecke, Physica A 246 (1997) 519. 
 
\bibitem{MH2} M. Hennecke, Physica A 252 (1998) 173. 
 
\bibitem{stauffer} D. Stauffer, J. Phys. A 27 (1994) 5029. 
 
\bibitem{maj} S. N. Majumdar, A. J. Bray, S. J. Cornell, C. Sire, Phys. Rev. Lett. 77 (1996) 3704. 
 
\bibitem{derrida4}A. J. Bray, B. Derrida, C. Godreche, Europhys. lett. 27 (1994) 177. 
 
\bibitem{derrida5}B. Derrida, J. Phys. A: Math. Gen. 98 (1995) 1481. 
 
\bibitem{jain} S. Jain, Phys. Rev. E 59 (1999) R2459. 
 
\bibitem{cm} C. M. Newman, D. L. Stein, Phys. Rev. Lett. 82 (1999) 3944. 
 
\bibitem{jain1} S. Jain, Phys. Rev. E 60 (1999) R2445. 
 
\bibitem{jain2} S. Jain and T. Yamano, Int. Jour. of Mod. Phys. C 19,(1) (2008) 161. 
 
\bibitem{ba} R. Albert and A.L. Barab\'asi, Rev. Mod. Phys. 74, 47 
 (2002). 
 
\bibitem{erdo} P. Erd\"os and A. R\'enyi, Publ. Math.  Debrecen 6, 290 (1959); P. Erd\"os and A. R\'enyi, Publ. Math. Inst. Hung. Acad. Sci. 5 , 17 (1960); P. Erd\"os and A. R\'enyi, Bull. Inst. Int. Stat. 38 , 343 (1961). 
 
 
\bibitem{howard}C. D. Howard, J. Appl. Prob. 37 (3) (200) 736. 
 
\bibitem{lima} F.W.S. Lima, R. N. Costa Filho, U.M.S. Costa, Journal of Magnetism and Magnetic Materials 270, (2004), 182. 
 
\end{thebibliography}
\end{document}